# Force-noise spectroscopy by tunnelling current deflection sensing

Markus Herz and Elke Scheer*

Department of Physics, University of Konstanz, D-78457 Konstanz, Germany

**An electro-mechanical setup for the measurement of AC-forces in a low-temperature tunnelling microscope has been developed, which enables extremely high force resolution. The crosstalk of vibrations onto the tunnelling current is used to measure the deflection of a force-sensing cantilever beam. We demonstrate its capability to measure the noise of the force at a tunnelling contact using polycrystalline Iridium. Depending on temperature, spring constant and current, a resolution in the range of fN/√Hz is possible. We observe peak levels of the force-noise at the energy of the expected phonon maximal density of states, which suggests that inelastic transport processes contribute to force fluctuations.**

___________

*Electronic address: elke.scheer@uni-konstanz.de*

# 1    Introduction

Current-induced forces can be strong enough to give rise to large effects in the electrostatic force between tip and sample of a scanning tunnelling microscope (STM), by changing the local resistance due to scattering of carriers,[1] and to enable the reversible switching of the atomic configurations in nano-systems.[2-4] The mechanism of these atomic rearrangements is an active field of research[5-7] for fundamental reasons as well as because these effects give rise to failure of nanoelectric circuits. Although not clarified in detail it is believed that subthreshold current densities activate position fluctuations the amplitude of which is finally sufficient to trigger the rearrangement.[4-7] However, this cannot straightforwardly be proven experimentally, because neither the current nor the shot noise is sensitive to fluctuations in the regime of tunnelling of independent charge carriers.[8] However, as we argue here, the simultaneous measurement of the current-induced forces and the charge current can give valuable information about the charge transport mechanism and the interaction of the charge carriers with their environment. Although technically relatively simple, measuring the naturally given noise of the current-induced forces in a combined tunnelling and force microscope does not seem to be an established method to date. This is the starting point of our approach, in which we detect the crosstalk of the exerted forces of the tunnelling electrons in the tunnel current. The force fluctuations carry the information about the correlations of the tunnelling electrons and therefore about the interaction mechanisms causing the forces.[5]

When a persistent flow of electrons is measured across a vacuum gap between metals, the averaged DC current changes by roughly one order of magnitude for a displacement of ~100 pm, usually normal to the surface, and in a nearly exponential way with respect to the displacement.[9] For a corresponding tunnelling current slope of 1 nA/nm and a measurement resolution in the range of 1 fA, it follows directly that by sensing changes of the tunnelling current, a displacement measurement resolution in the range of 1 fm is possible. Although it was proposed to use a tunnelling gap for measuring the deflection of a cantilever beam in the initial times of the Atomic Force Microscope (AFM),[10] and successful attempts to use this technique have been performed,[11] to the best of our knowledge it has later not been used extensively because of its apparent complexity.

Here, we exploit the approach to replace the independent deflection sensing of a conductive cantilever housing the tip of a STM by carefully calibrating the signals obtained from the cross-talk of the deflection on the tunnelling current. The cross-talk essentially depends on the spatial current derivative in the direction of the observed cantilever vibration, which is calibrated by an off resonance measurement at low frequency. While constituting one of the most sensitive measurement devices in the range of acoustic frequencies, the tunnelling gap can simultaneously be used to measure dynamic displacements with outstanding resolution.

# 2    Force noise measurements

## 2.1    Design of the experiment

We measure the oscillation of a conductive cantilever beam in the field of the tunnelling current distribution depending on the relative position between the two electrodes in the tunnelling microscope. The direction of oscillation is normal to the surface.

**Figure 1** shows the experimental setup and detailed views of cantilever with sample, as well as the samples mounted in the STM. The eigenfrequency of the vertical oscillation of the top wire-hook has been estimated to be a few kHz above the cantilever eigenfrequency that amounts to approximately 2.7 kHz, depending on the contact configuration. The cantilever supports a sharp blade of the metal under investigation (here: Iridium) which faces another sharp blade (of Ir) on the opposite side of the junction. Both blades were cleaved and then electro-polished, using a high-purity polycrystalline Ir-wire. These two electrodes are arranged perpendicular to each other and are therefore referred to as "cross-bar" sample. The electro-polishing was performed for few seconds with a saturated $CaCl_2$ solution according to ref. [12], AC voltages of up to 9 V, and frequencies in the kHz range.

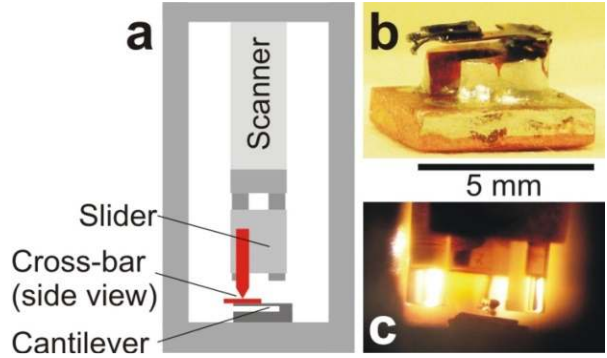

**Fig. 1:** Experimental setup and detailed views. **a** Schematic diagram of the STM, carrying the cantilever mounted on the sample holder, and the cross-bar contact, with one of the bars mounted on the cantilever, the other on the slider. **b** Detailed view of the conductive cantilever mounted on the sample holder, with the Ir blade mounted on it. **c** Detailed view of the cross-bar configuration before the STM is evacuated for cool-down. The scale bar refers to images **b**, **c**.

The principle of the force-fluctuation measurement relies on detecting the changes of the spectral content of the peak, i.e. simply the AC current caused by the resonance-enhanced cantilever oscillation, caused by the force fluctuations, in the tunnelling current. The measurement aims at the determination of the mean square displacement $\langle z_c^2 \rangle$ of the cantilever at the position of the tip, namely

$$\langle z_c^2 \rangle = \frac{\langle I_{AC,c}^2 \rangle}{\left(\frac{\partial I_t}{\partial z}\right)^2} \qquad (1).$$

This quantity depends on the measurement of the cantilever deflection-induced mean square current fluctuation $\langle I_{AC,c}^2 \rangle$ and the vertical gradient of the DC tunnelling current, $|\partial I_t/\partial z|$.

The proposed measurement scheme simplifies the requirements for the measurement set-up, because no special circuit for the force measurement is necessary, and no Phase-Locked Loop circuit is necessary either. The existence of a tunnelling contact is verified by observing an exponential current-distance curve.[9] During measurements in the low-temperature STM, we ramp the voltage in feedback-off mode of the STM once a tunnelling contact has been established, and measure the spectral content of the cantilever oscillation, $\langle I_{AC,c}^2 \rangle$, usually by heterodyne coupling. The detection can alternatively be performed by using a spectrum analyser and post-processing of the obtained power spectrum. During spectroscopy, we simultaneously observe the mean frequency of the oscillation peak to stay in the used measurement band for assurance of the detection stability.

For calibrating the voltage-dependence of the cross-talk measurements at constant conductance, a separate voltage ramping is performed before and after the main measurements, where we deliberately oscillate the tip at frequencies slightly below 200 Hz, far off resonance, and with small amplitudes in the range of few pico-metres, see supporting information. An exemplary $|\partial I_t/\partial z|$ calibration curve can be seen in **Fig. 3b,** below. Repeatedly we measure the noise and calibrate this scaling, to exclude any significant tip changes. We additionally determine changes of the noise level due to the influence of additional external voltage noise on the deflection noise in additional measurements at the same conductance, to exclude possible artefacts due to tip-sample capacitance gradients, and estimate possible capacitive readout errors, see supporting information. How we calibrate the obtained AC-current quantitatively, and how the background noise is subtracted, is described in more detail in the Electronic Supplementary Information (ESI).

Further in the ESI, a rationale for the selection of quality factor, frequency and spring constant of the cantilever for the case of an unperturbed oscillation is given. We also include a breakdown of relevant error contributions to the measurement, a detailed description of the processing scheme to obtain the mean square deflection of the cantilever including calibration procedures. A verification procedure allows excluding an additional influence of external voltage noise on the deflection and at the same time of a capacitive current on the deflection measurement.

**Figure 2** shows a topographic image obtained in STM mode with the cross-bar configuration shown in **Fig. 1**. In the centre of the image, the tip was stabilized and spectroscopy was performed.

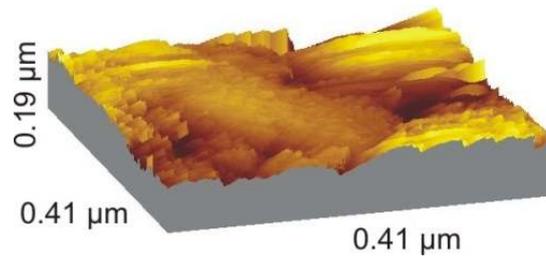

**Fig. 2:** Topographic image obtained in STM mode with the cross-bar configuration shown in Fig. 1. Close to the centre of the image, the tip was stabilized and spectroscopy was performed.

## 2.2  Calibration procedure and results

**Figure 3a** shows a measurement of the current-voltage characteristics of an Ir tunnelling contact of tunnelling resistance $R_t \approx 100$ MΩ, **b** the calibration measurement $|\partial I_t/\partial z|$, **c** raw data of a RMS-amplitude measurement with the envelope roughly scaling with $|\partial I_t/\partial z|$, i.e. $\langle I_{AC,c}^2 \rangle^{1/2}$ including small offsets that are later removed, see ESI, and **d** the final calculated energy ratio $E_r = k_c \langle z_c^2 \rangle / k_B T$, where $k_c$ is the effective spring constant and $z_c$ is the displacement of the tip.

It can be seen that the nearly linear current-voltage characteristic also produces a very similar, nearly linear response in $\partial I_t/\partial z$, in accordance with the exponential distance dependence expected for tunnelling through a barrier $I_t = I_t(V)e^{-2\kappa z}$, where $V$ is the tunnelling voltage and κ is the decay constant of the tunnelling current. The residual oscillations are due to low-frequency modes of the mechanical dampers in the STM setup, which are roughly of constant amplitude and consequently also produce a cross-talk amplitude linear in the current. The root–mean-square (RMS) amplitude envelope, **Fig. 3c**, shows that the cantilever fluctuation is also roughly determined by the linear current dependence of the crosstalk, a proving the measurement principle works. The calculated energy ratio $E_r$ is the quantity of highest interest, since thermodynamic equilibrium would produce a constant $E_r = 1$, and the deviations are due to the physics of the contact.

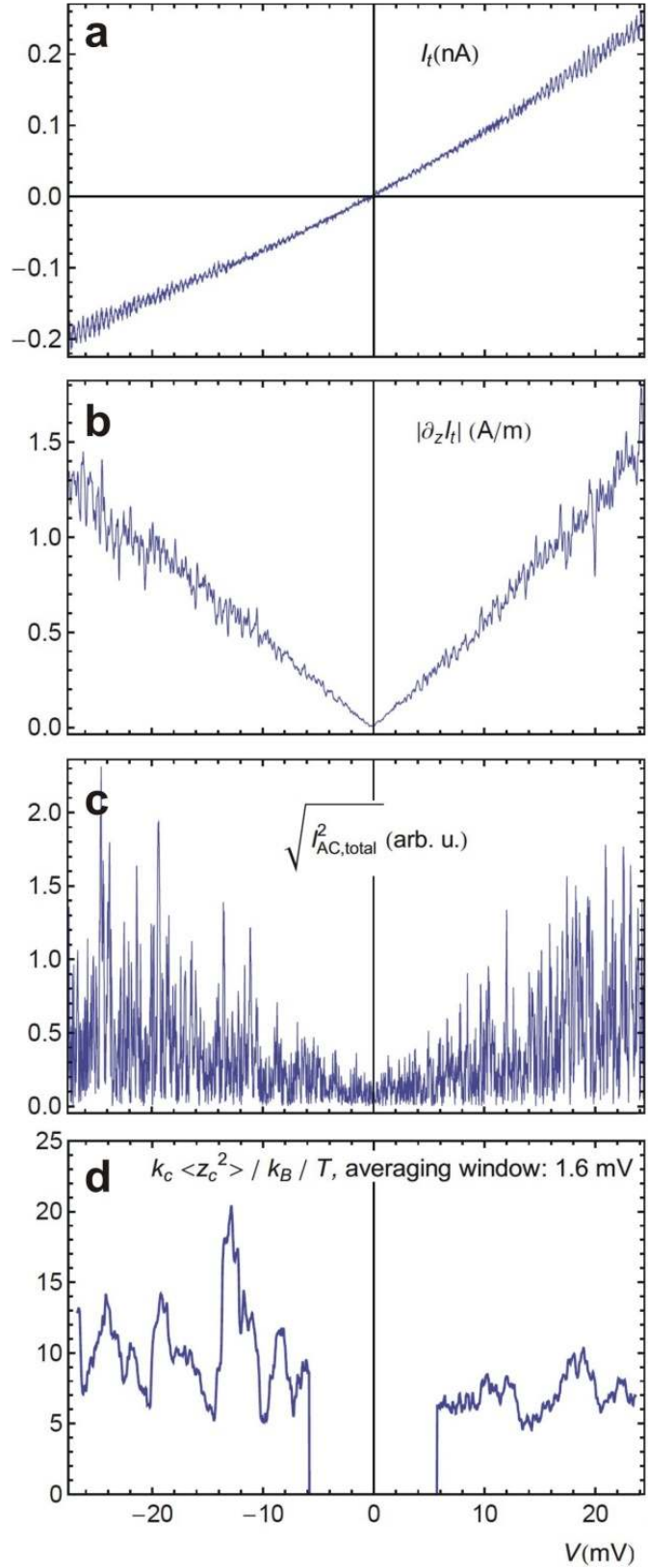

**Fig. 3 a** Measured current-voltage characteristic of an Ir tunnelling contact. **b** Current gradient $|\partial I_t/\partial z|$ measurement used for calibration, **c** Raw data of RMS-amplitude measurement with an envelope roughly scaling with $|\partial I_t/\partial z|$ single scan. **d** Final calculated energy ratio of the cantilever oscillation, using the sum of four voltage scans. Single scan duration: 52s. Data areas with high relative inaccuracy due to the smaller signal for $|V| \leq 5.8$ mV have been replaced by zero. $T = 5.5$ K.

The external voltage noise input used for two of the four sweeps, which is larger than the amplifier noise, and which is applied on the non-amplifier side of the contact, did not result in enhancements of the measured deflection noise, excluding any artefacts of the tip-sample capacitance gradients. Therefore we show in **Fig. 3d** the averaged residual deflection noise that was calculated from the sum of four consecutive voltage sweeps and the individual residual deflection noises. The scheme for processing the data of the cantilever deflection measurements is presented in the supporting information section. Averaging over 50 data points (1.6 mV) has been performed with the final processed data. The shot-noise showed a behaviour typical for single electron tunnelling with a Fano Factor $F=1$, see ESI and ref. [8]. Offset shifts between different sweeps are so small that possible residual error contributions due to mechanical drift variations, which would be corrected for in first order as described in the supporting information, are negligible.

The energy ratio $E_r$ shows that the cantilever energy is above the thermal energy by up to an order of magnitude, or an amplitude ratio of up to three. The tunnelling contact thus determines the dynamics of the cantilever. This can also be derived from the analysis of the distance dependence of the tunnelling gap that determines the quality factor of the cantilever as can be seen by the rapid oscillation of the data in **Fig. 3c**. The freely oscillating cantilever has a quality factor above $10^4$ that would result in a much slower variation on the scale of tens of seconds, corresponding to approximately five periods per voltage scan. The quality factor of the free cantilever was determined by a capacitive current ring down measurement. Since the quality factor is determined by the tunnelling contact, the damping of the oscillator is also determined by this contact. Accordingly, the average energy of oscillation will be determined by an effective temperature of this contact. Thus we can understand the slight asymmetry of the energy ratio, with respect to voltage, by different cantilever heating at different polarities. The voltage-symmetric peaks at $V \approx \pm 18$ mV are comparable to the maximum of the phonon density of states of Ir.[13] Interestingly, even the asymmetric shape of the calculated phonon density of states main peak[13] feature is mimicked by these peaks. The symmetric peak at $V \approx \pm 12$ mV might be determined by surface properties.

## 3    Discussion

It is well known that inelastic transport properties in off-resonant tunnelling regime enhance the tunnelling probability. This fact is used as a tool to determine the phonon density of states by inelastic electron tunnelling spectroscopy (IETS):[14] At voltages corresponding to phonon energies a kink in the current is observed and the current-voltage characteristic features an enhanced slope above this threshold. The constant increase of the conductance signals that above this threshold phonon excitation is possible with voltage-independent probability. In analogy to IETS it has been shown theoretically that for very small tunnelling probability also the shot noise rises abruptly above the voltage that corresponds to the excitation of the phonons $eV \geq \hbar\omega_{ph}$.[15,16] When increasing the tunnelling probability the contribution of the inelastic transport properties to the shot noise may also be negative,[15-17] but in any case, kinks of the shot noise at these voltages are expected. At first glimpse one could expect the same behaviour for the force noise, i.e. kinks at voltages corresponding to phonon energies. However, taking into account the physical processes giving rise to force fluctuations, the assumption of constant probability above a threshold value is not fulfilled. First theoretical investigations treating current-induced forces in nanocontacts establish a correlation between the Nyquist and shot noise and the Langevin force fluctuations.[5,18] However, this correlation is more complex, because of the appearance of non-conservative forces and nonlinear dynamics due to the coupling of vibrational modes to electronic modes. The force fluctuations couple to the electronic charge, what is most pronounced at the Fermi edge where screening is less efficient. Experimentally we do not observe kinks but peaks at the voltages corresponding to the excitation of bulk phonon energies of Ir.[13] This suggests that the force-noise signal might be proportional to the phonon density of states. A tentative explanation for the appearance of peaks of the force-noise signal is as follows: The exertion of a force is a dynamic effect involving redistributing the occupation of electronic states in both electrodes. Thus it can be expected to be most effective at the Fermi energy, as described by the theory of dynamical Coulomb blockade.[19] For the same reason also the variations of the force are most pronounced at the Fermi edge where the distribution function deviates from zero or unity, respectively. Hence, while the charge transport itself is still tunnelling of independent carriers, their dynamics are correlated and their correlation is most pronounced at the Fermi edge. Summarizing, we can expect the effect to be proportional to the phonon density of state $Z(\omega)$, the electron phonon coupling constant $\lambda$, and the derivative of the Fermi function with respect to the energy $\boldsymbol{E_r \propto Z(\omega)\, \lambda\, \partial f/\partial E}$.[18,19]

## 4      Conclusions

In conclusion, we have presented a measurement setup and a processing scheme for measuring the deflection of a macroscopic cantilever in the tunnelling microscope by observing the linear cross-talk of the deflection in the tunnelling current. The deflection-background spectral density is around 10 fm/√Hz. This goes beyond high-resolution AFM methods, using cooled amplifiers close to the sample, by around one order of magnitude in the power spectral density of the background deflection noise.[20] Today, there are high frequency readout techniques for determining the position noise of nanofabricated cantilevers using coupling to single-electron transistors[21,22] or, similar to our method, coupling to atomic point contacts.[23] In these techniques, higher resolution compared to our method would be possible, and back-action forces have already been observed,[22,24] displaying also coupling to mechanical modes of a macroscopic cantilever.[24] However, these techniques involve demanding nanofabrication, and to our knowledge an application of these building blocks in STM has not been reported. The new technique allows for enhanced resolution of the measured deflection in general, and for the measurement of force induced fluctuations caused by the dynamics of the system in non-equilibrium. We find that the voltage-symmetric energy corresponding to the peaks of the observed force noise coincides with the energy of a maximum of the phonon density of states of crystalline Ir, revealing the importance of vibrational excitations in the environment of the contact for the force exerted on current-carrying charge carriers. Future investigations will aim at the disentanglement of the various measured contributions, and how they depend on material properties like the conductance of the electrode material. It is conceivable that investigations on a great variety of systems could reveal electronic correlations in the transport also in the far tunnelling regime. These expected results wold be most valuable for understanding the underlying physics of transport in both nano-systems and bulk materials, where scattering, fluctuations and local excitation play an important role.

**Acknowledgements**

For illuminating discussions regarding the physical concepts of force noise the authors wish to thank W. Belzig, P. Leiderer, F. Pauly, G. Rastelli, C. Cuevas, and B. Reulet. We are grateful to T. Pietsch, M. Fonin, S. Diesch, R. Sieber, and H. Zheng, for experimental advice and assistance. We thank the members of the technical services of the university for technical support. Financial support from the DFG through SFB767 "Controlled Nanosystems" is gratefully acknowledged.

# Electronic Supplementary Information for manuscript
# "*Force-noise spectroscopy by tunnelling current deflection sensing*"

*Markus Herz and Elke Scheer*

Department of Physics, University of Konstanz, D-78457 Konstanz, Germany

1. Selection of spring constant, quality factor and frequency of the unperturbed oscillation
2. Breakdown of relevant error contributions of the measurement
3. Measurement setup
4. Verification procedure for excluding influences of external voltage noise on the measurement
5. Processing scheme for cantilever RMS deflection measurement including calibration procedure
6. Expected deflection measurements for negligible forces from the contact
7. References

---

1. **Selection of spring constant, quality factor and frequency of the unperturbed oscillation**

In this section we give a rationale for the selection of the dynamic properties of the mechanical oscillator in use.

*Spring constant*

Force noise spectroscopy is proposed to concentrate on the determination of the deflection $z_t$ of a cantilever in the force field of a possibly microscopic contact that acts in addition to the cantilever's restoring force $k_c\, z$, where $k_c$ the spring constant, and $z$ the displacement. The relation between a force-noise power spectral density $S_F$, sometimes called noise drive [S1] and the dynamic mean square displacement of the mounted cantilever is given by

$$\langle z_t^2 \rangle = \frac{S_F Q \omega_0}{4 k_c^2}. \tag{1}$$

Here, $Q$ is the quality factor of the oscillation and $\omega_0 = 2\pi f_0$ the angular eigenfrequency of the oscillator. According to eq. (1), it would be easy to maximize the response of any force-noise in the displacement by minimizing the spring constant. Today it is well known that there are lower limits for the spring constant in order to allow stable cantilever operation without the need of using large amplitudes of oscillation to overcome the Jump-To-Contact issue [S2]. When $k_c$ reaches the order of magnitude of a single chemical bond stiffness, the contact can easily be closed by a jump of the cantilever enabling a cluster of atoms or a molecule to impede the free oscillation in the scalar field of the tunnel current. To be able to observe dynamic forces close to the contact, we follow these findings and use a cantilever with a spring constant larger than that of a chemical bond by one order of magnitude. It is later shown that with $k_c \approx 1$ kN/m, also the calibration procedure is simplified.

During the development of the setup it became also clear that the cantilever can be prone to electric self-oscillation, probably caused by hysteretic discharge. The oscillation is then accompanied by repeated hard crashing of the contact, increasing the area of contact and impeding further stable operation. We anticipate that despite of these known issues, it is possible to observe forces at lower spring constants, and modify the later calibration procedures according to the frequency shift that is related to an additional tip-sample spring constant to be taken into account.

*Quality factor*

According to eq. (1), the quality factor should be maximized for maximum response. We anticipate that the quality factor can be determined by the tip-sample contact which is then the natural limit.

The cantilever was fabricated from polycrystalline phosphorus bronze known to exhibit extremely high quality factors at low temperatures, and polished. In our measurements, we observed different quality factors from different tip-sample interactions. With the observation of a decaying capacitive current oscillation for tens of seconds on the oscilloscope, we estimate the cantilever quality factor to be in the range of at least $10^4$ in vacuum and at low temperatures. The experimental quality factor in operation is typically in the range of $10^3$ or even below. This shows that the energy loss is usually not determined by the cantilever, and the response is maximized according to the contact's contribution to the energy loss.

*Frequency*

Another accessible dynamic property is the frequency of the fundamental oscillator mode. According to equation (1), it would be desirable to maximize this frequency. However, the background noise due to the amplifier's contribution and capacitive charging of the input capacitance using a trans-impedance amplifier also increases with increasing frequency. We propose that this detector noise should not significantly reduce the measurement quality.

In equilibrium, i.e. without current flowing across the junction, we expect a thermal motion of the cantilever. According to the equipartition theorem, there is the thermal energy $k_B T$ of the oscillation, including potential and kinetic contributions and the theorem requires

$$\left\langle z_{t,thermal}^2 \right\rangle = \frac{k_B T}{k_c} \tag{2}$$

Accordingly, the measurements can be compared to the thermal contribution, and will be determined by its noise, if the thermal contribution is present. According to the above, the amplifier noise, inside a detection bandwidth, should not exceed this thermal equilibrium contribution. Since the amplifier noise is increasing with increasing frequency, this equality is typically fulfilled for

$$\frac{k_B T}{k_c} \left(\frac{\partial I_t}{\partial z}\right)^2 \approx S_{amp}(f) \frac{\omega_0}{2\pi\,100} \tag{3}$$

The calibration factor $\partial I_t / \partial z$ is squared and included in eq. (3) for transfer into the regime of square current fluctuations. Its acquisition is described later in the description of the processing. According to eq. (3), the current fluctuations caused by the cantilever resonance, and its equilibrium thermal content, is selected to be equal to the amplifier's noise background in a bandwidth that amounts to 1/100 of the fundamental frequency, see factor 100 in eq. (3). Consequently, for $Q = 100$, roughly all of the resonance's spectral content would be included in this measurement band. Due to this proposed equality, the background noise would not reduce the measurement resolution at the same time. Since first experimental tests show that the quality factor is usually one order of magnitude higher, the measurement is usually well defined by this selection of the measurement bandwidth, without excessive background noise. Narrower bandwidths tend to be critical because of undefined influence of their limits on the captured spectral content and possibly drifting resonance frequencies during measurement.

Since the typical quality factor of the oscillation is much higher than 100, it is possible that the detector noise can still be reduced in comparison with the detected resonance curve's spectral content. This can be realized by a Fourier transformation of the detected signal, and concentration on the peak. However, also when taking the full noise power inside the detection bandwidth, according to eq. (3), the background noise will just equal the thermal noise level, namely the thermal equilibrium part of the resonance curve's spectral content. Possible shot-noise contributions will be discussed later. These uncorrelated noise offsets of comparable size are usually not of major concern.

Following these considerations, depending on the temperature, for a low-temperature STM operating between 0.3 K and 5 K, and a current of $I_t = 100$ pA, the optimal frequency is in the range of few kHz. In principle the optimization and mechanical setup/frequency adaptation would need to be repeated for each and every current value to be observed. However, starting with a typical low current value is selected as a first choice, since frequent rearrangements of the possibly contaminated tunnelling contact at higher currents, closer to the mechanical contact, may make the measurement difficult. Stability over time in the range of several minutes or hours is required. Moreover, the force noise can already be observed at higher currents, and the resolution would likely be better due to increased force-noise by itself. On the other hand, to impede possible observation of the force-noise in the middle of useful STM currents, by optimizing the setup for very high current only, seemed to be not the first choice, although it might be considered later. We have selected $f_0 = 2.7$ kHz, in order to be prepared for operation at lower temperatures, to easily find the resonance line, and to minimize the above mentioned background errors.

It should be noted that there is a systematic error, because the resonance curve has contributions at all frequencies. This error, usually small, should be estimated and corrected or discussed, depending on the measurement goals. E.g., for a fundamental frequency of 2.7 kHz and a bandwidth of 27 Hz, the difference can be modelled to be 6%. It can be reduced by using a higher bandwidth. By using lower frequencies, we also minimize capacitive currents that might give an error contribution to the oscillation readout, and that are discussed in section "Verification procedure for excluding influences of external voltage noise on the measurement".

## 2. Breakdown of relevant error contributions of the measurement

Here, we discuss a selection of presumably significant error contributions (besides the thermal noise discussed above) and how we bring them to a level that is below the detection noise determined in the preceding section.

*Mechanical-External Noise*

According to equations (1) and (2), and depending on temperature, the achievable noise floor of the tunnelling detector is in the range of several $10^{-14}$ N/Hz$^{1/2}$. Simple estimations of the displacement noise spectral density at the selected frequency and at the base of the cantilever are in the range of $10^{-16}$ m/Hz$^{1/2}$ for a very quiet room, this would result in $10^{-13}$ N/Hz$^{1/2}$ acting on the oscillator. Although the experiment was performed in a low-noise lab, we could not determine if a stability of a few am/Hz$^{1/2}$ is reached, that would be required to make the measurement independent of vibrations. However, we could easily solve this question by another vibration issue at low frequencies, which is typically present in STM and AFM, where vibration modes, e.g. of the cryostat, cause considerable deflections inside the microscope that degrade the measurement. This issue made it necessary to implement dual mechanical vibration insulation in vacuum, and with resonance frequencies of a few tens of Hz. Making use of this vibration insulation, also at higher frequencies, we expect to be able to reduce the externally caused vibration level at the measurement device to a negligible level in the range of 1 am/Hz$^{1/2}$.

*Mechanical Internal Noise*

Including the electronic piezo filters described with the measurement setup [S3,S4], a stability of $10^{-17}$ m/Hz$^{1/2}$ can be reached at the scanning piezo. It could be further suppressed by more than one order of magnitude by placing the cantilever on the usual side of the sample holder, instead of on the piezo. In this way, this noise contribution is supressed sufficiently.

*Electrical Noise*

Possible electrical force noises and readout noises are discussed in the section "*Verification procedure for excluding influences of external voltage noise on the measurement*"

### 3. Measurement setup

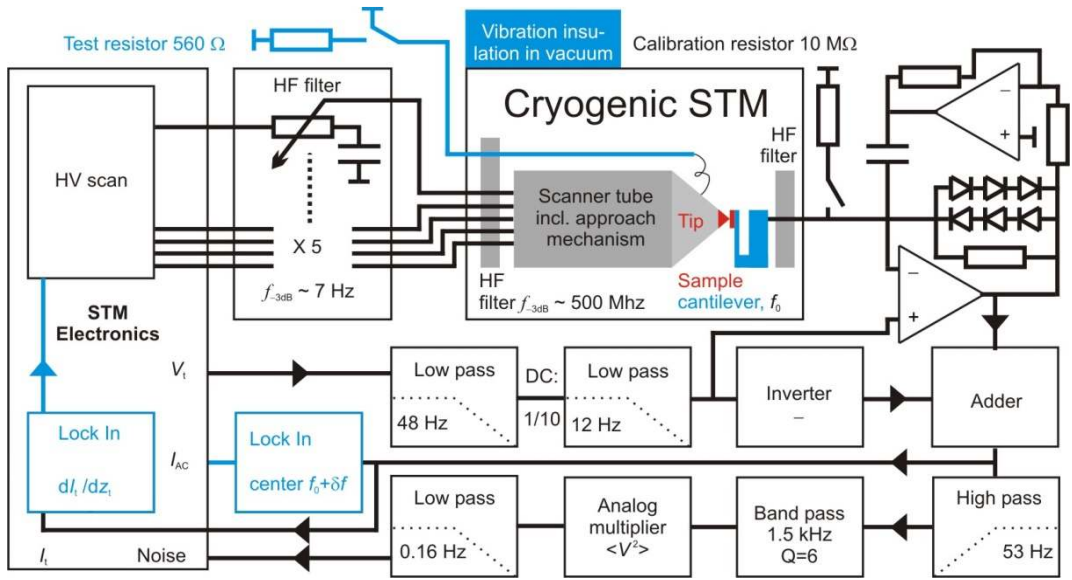

**Figure S1**: Schematic diagram of the measurement setup.

A schematic diagram of the measurement setup is shown in Fig. S1. It is very similar to the setup used in ref. [S3]. The blue and red components mark significant functional and geometric changes. The noise of the amplifier has been significantly reduced. The test resistor allows for applying a defined noise level needed for the noise immunity verification procedure described in the following section. Dual stage passive vibration insulation was applied in vacuum to reach the lowest possible vibration levels in compliance with the findings presented in the previous section. Two lock-in detectors allow for

- Determination of cross talk of displacement to current readout, calibration $\frac{\partial I_t}{\partial z}$.
- Heterodyning of mechanical oscillator vibration line (current signal) with lock-in oscillator pre-selected frequency $f_0 + \delta f$.

## 4. Verification procedure for excluding influences of external voltage noise on the measurement

The measurement setup includes the input of a defined voltage noise at the junction. This allows for estimating influences of tip-sample capacitive gradients in the tip-sample contact and their coupling with the

- Voltage noise, causing AC forces.
- DC voltage, causing capacitive currents and possible readout errors.

The electrical forces due to tip-sample capacitance $C_{ts}$ gradients in the direction of the oscillation, Voltage $V_{ts}$ and its noise spectral density $\delta V_{ts}$ can be calculated according to

$$S_{F,el} = \left(\frac{\partial C_{ts}}{\partial z} V_{ts} \delta V_{ts}\right)^2. \tag{4}$$

By feeding an additional voltage noise $\delta V_{ts}$ to the junction, it is observed if the measured noise level caused by the cantilever oscillation would change. If not, obviously this contribution is negligible. This observation allows for estimating the influence of capacitive currents due to charging of the tip-sample capacitance. If the electrical contribution shown in equation (4) is much smaller than the thermal, the following relation is found from (1), (2) and (4)

$$\frac{\partial C_{ts}}{\partial z} \ll \frac{\sqrt{\frac{2 k_B k_c T}{\pi f Q}}}{V_{ts,0} \delta V_{ts}} \tag{5},$$

which gives an estimation for an upper limit of the tip-sample capacitance gradient in the direction of the oscillation. Then, we find for the mean square current caused by capacitive effects and thermal motion

$$\langle \delta I_{C_{ts}}{}^2 \rangle = \left(2\pi f V_{ts} \frac{\partial C_{ts}}{\partial z}\right)^2 \frac{k_B T}{k_c} \tag{6}.$$

On the other hand, the unavoidable thermal contribution of the current fluctuation is

$$\langle \delta I_{thermal}{}^2 \rangle = \frac{k_B T}{k_c} \left(\frac{\partial I_t}{\partial z}\right)^2 \tag{7}.$$

Combining (6) and (7), and using the assumption made in (5), we find

$$\frac{\langle \delta I_{C_{ts}}{}^2 \rangle}{\langle \delta I_{thermal}{}^2 \rangle} = \left(2\pi f V_{ts} \frac{\frac{\partial C_{ts}}{\partial z}}{\frac{\partial I_t}{\partial z}}\right)^2 \ll \frac{8\pi f k_B k_c T V_{ts}{}^2}{\delta V_{ts}{}^2 Q V_{ts,0}{}^2 \left(\frac{\partial I_t}{\partial z}\right)^2} \tag{8}.$$

For typical values of $\delta V_{ts}$=3.5 nV/Hz$^{1/2}$, $f$=2.7 kHz, $Q$=500, $\frac{\partial |I_t|}{\partial |z_t|}=\frac{V_{ts}}{V_{ts,0}}$A/m, $k$=1kN/m, we find

$$\frac{\langle \delta I_{C_{ts}}{}^2 \rangle}{\langle \delta I_{thermal}{}^2 \rangle} \ll \begin{cases} 0.84 & T = 5.5\text{K} \\ 0.05 & T = 0.3\text{K} \end{cases} \tag{9}.$$

Thus, we confirm that the contributions of capacitive currents are negligible, as long as no change in the measurement is detectable due to the external noise-input $\delta V_{ts}$. Larger external noise results in a possibly smaller upper limit for the tip-sample capacitance. However, it may increase the probability of exceeding the intrinsic force noise.

## 5. Description of the processing scheme for cantilever RMS deflection measurement including calibration procedure

The measurement aims at the determination of the mean square displacement of the cantilever at the position of the tip,

$$\langle z_c^2 \rangle = \frac{\langle I_{AC,c}^2 \rangle}{\left(\frac{\partial I_t}{\partial z}\right)^2} \qquad (10).$$

This quantity depends on the measurement of the cantilever deflection-induced mean square current fluctuation $\langle I_{AC,c}^2 \rangle$ and the vertical gradient of the DC tunnelling current, $\partial I_t / \partial z$. The quantity $I_{AC,c}$ is the deflection signal of the cantilever vibration in the current regime. It is determined by a lock-in measurement -with an exemplary transfer function shown in Figure S2.

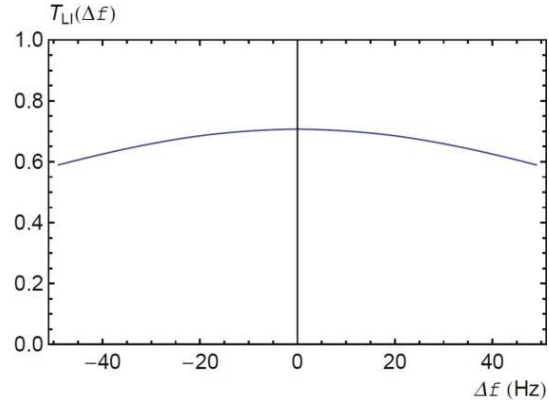

**Figure S2**. Exemplary lock-in transfer function $T_{LI}(\Delta f)$ in the relevant frequency band, integration time: 1 ms. The transfer function is unity just exactly at $\Delta f = 0$, which is not displayed here, and which has usually no impact on the measurement.

The frequency shift $\Delta f$ is the deviation of the measured noise frequency from the lock-in oscillator frequency at which the incoming noise is mixed in the lock-in. The transfer function $T_{LI}(\Delta f)$ allows the determination of $I_{AC,c}$. However, the offset noise (tip retracted) and an estimation of the shot-noise have to be subtracted from the heterodyne noise output $I_{AC,total}^2$ before, in order to obtain the current component $I_{AC,c}^2$ that belongs only to the cantilever vibration. We thus obtain

$$\langle z_c^2 \rangle = \frac{I_{AC,total}^2 - \langle I_0^2 \rangle - \frac{S_{measured}}{S_{amplifier}}\langle I_0^2 \rangle}{T_{LI}^2(\Delta f)\left(\frac{\partial I_t}{\partial z}\right)^2} \qquad (11).$$

The quantity $\langle I_0^2 \rangle$ is the lock-in filtered offset current fluctuation mean square amplitude of the amplifier, when tip and sample are not in vicinity. It belongs to a Power Spectral Density (PSD) $S_{amplifier}$. The quantity $S_{measured}$ is the measured current noise PSD. If the noise from the contact is determined by shot noise $S_{measured} = 2eI_t$ typical for single electron tunnelling with a Fano Factor $F = 1$, the calibration can be simplified by using this fitting approximation.

The final required calibration input is the vertical gradient component, $|\partial I_t / \partial z|$. It is determined by another lock-in measurement at ~5-10% of the cantilever resonance frequency, to avoid significant resonance overshooting for the calibration. The measurement $|\partial I_t / \partial z|$ is usually observed to be linear in the current, and can be fitted to the $I_t$ curve, to find a scaling factor used for the compensation of possible small variations of the current due to drift variations and noise, compared to the calibration sweep.

It is assumed that all measurements are calibrated according to the transfer functions of the amplifier.

We note that it is required to observe the absolute frequency of resonance, since its shift gives information about possibly significant vertical force gradients compared to the cantilever spring constant, that could give rise to additional error contributions of the denominator of $|\partial I_t/\partial z|$. An error discussion for the folding of the resonance curve with the lock-in transfer function has already been given in section 1 of this document.

## 6. Expected deflection measurements for negligible forces from the contact

For negligible forces from the contact, e.g. when the thermal energy of the cantilever would produce a constant average RMS displacement, the RMS deflection crosstalk in the current is expected to be linear in the vertical gradient component, $|\partial I_t/\partial z|$, which is usually linear in the voltage, for small voltages.